
\input epsf
\magnification=\magstep1
\overfullrule=0pt
\baselineskip=18 truept
\font\mein=cmbx10 at 14.4 truept
\centerline{\mein Search for the Hypothetical $\bf \pi \to \mu x$
Decay$^\ast$}
\medskip
\centerline{R. Bilger$^a$, H. Clement$^a$, A.
Denig$^b$, K. F\"ohl$^a$, P. Hautle$^c$,  W. Kluge$^b$,
J. A. Konter$^c$,}
\centerline{G. Kurz$^a$, S. Mango$^c$, D. Schapler$^a$,
F. Sch\"onleber$^b$, U. Siodlaczek$^a$,}
\centerline{B. van den Brandt$^c$, G.J. Wagner$^a$, R. Wieser$^b$}
\medskip
\centerline{$^a$Physikalisches Institut, Universit\"at T\"ubingen,
D-72076 T\"ubingen, Germany}
\centerline{$^b$Institut f\"ur Experimentelle Kernphysik,
Universit\"at Karlsruhe, D-76021 Karlsruhe, Germany}
\centerline{$^c$Paul Scherrer Institut, CH-5232 Villigen,
Switzerland}
\bigskip
\noindent {\bf Abstract:} The KARMEN collaboration has reported the
possible observation of
\footnote{}{$^\ast$This work has been supported by the German
Federal Minister for Education and  Research  (BMBF) under contract
numbers 06 T\"U 669 and 06 KA 564, and by the DFG (Mu 705/3,
Graduiertenkolleg)}  a hitherto unknown  neutral and
weakly interacting particle $x$, which is produced in the decay
$\pi^+ \to \mu^+ x$ with a mass $m_x = 33.9$ MeV. We have searched
for this hypothetical decay branch by studying muons from pion decay in
flight with the LEPS spectrometer at the $\pi E3$ channel at PSI and find
branching ratios $BR (\pi^- \to \mu^- \bar{x})
< 4 \times 10^{-7}$ and $BR (\pi^+ \to \mu^+ x) < 7 \times 10^{-8}$
(95\% C.L.).
Together with the limit $BR > 2 \times 10^{-8}$
derived in a recent theoretical  paper our  result would leave only
a narrow region for the existence of $x$ if it is a heavy neutrino.
\vfill\eject
\noindent {\bf 1. Introduction}
\medskip
\noindent The KARMEN collaboration, which studies nuclear
interactions of neutrinos resulting from the decay of stopped $\pi^+$
and $\mu^+$ in the beam dump at the Rutherford Appleton Laboratory
(RAL), recently reported [1] an anomaly in the time distribution of
single prong events concerning the time interval corresponding to
muon decay. The simultaneously observed energy deposit of these
anomalous events in the KARMEN detector corresponds to that of
neutrino interactions resulting in visible energies of typically
$T_{vis} = 11 - 35$ MeV~[1]. Hence this anomaly has been suggested to
originate from the observation of a hitherto unknown weakly
interacting neutral and massive particle, called $x$, which is
produced in the decay $\pi^+ \to \mu^+ x$ in the beam dump, then travels
with a velocity $v_x = c/60$ (corresponding to the anomaly at
3.6 $\mu$s in the time spectrum) and finally gets registered in the
detector volume after passing a steel wall of more than 7 m
thickness. The observed velocity and the two-body kinematics of the
assumed pion decay branch lead to a mass of the $x$-particle of $m_x
= 33.9$ MeV, which is just marginally below the kinematical limit and
leaves a
tiny reaction Q-value of 7 keV. Hence in the reference system of the
decaying pion the kinetic energies of the decay products are $T_x
\approx 5$ keV and $T_\mu \approx 2$ keV, respectively. A study of
this particular branch  by observing the
associately produced muons in the pion decay at rest is therefore
highly prohibitory. On the other hand the tiny Q-value is just ideal for
the observation of this decay branch in the pion decay in flight.
\bigskip
\noindent {\bf 2. The experiment}
\medskip
\noindent We exploit  the fact of a tiny Q-value to search for muons
from a possible	$\pi \to \mu x$  decay in flight, taking
advantage of the Lorentz boost. In a pion beam with momentum $p_\pi = 150$
MeV/c, a pion decay into $\mu$ and $x$ would result in a well-behaved
muon ``beam'' with momenta in the range of $p_{\mu} = 112.8 -
114.4$ MeV/c and a divergence of $\Delta \Theta_{\mu} \leq 5$ mrad. In
its chromatic mode the $\pi E3$ channel of the PSI delivers a pion
beam at the position where usually the target is installed (pivot
point), with a momentum resolution of $\Delta p/p \leq
0.3\%$ and a divergence  $\Delta\Theta < 40$ mrad FWHM [2]. The Low Energy
Pion Spectrometer (LEPS), a description of which is found in refs.
[3,4], has a momentum resolution of $\Delta p/p \leq 0.1\%$ and an
angular acceptance of 150 mrad. Hence the increase of divergence in
the $\mu$ ``beam'' from the decay $\pi \to \mu x$ is small compared to the
original beam divergence, whereas the increase in the momentum spread
caused by this decay should be sizeable and easily detectable by LEPS.
\medskip
\noindent The measurements have been carried out with two slightly
different setups. The setup of the first measurement is shown in
fig.~1, where LEPS is put ``in the beam'', i.e. at an angle of $0^\circ$.
The beam
is prepared in the Z-shaped $\pi E3$ channel containing two bending
magnets for momentum selection. Slits S 1 and S 2, set to
openings of 16 and 26 mm, respectively, in the central part of the
channel as well as a collimator (C) with a diameter  of 10 mm installed at
the last quadrupole magnet  of the beamline  ensure a
well-defined phase space of the beam, the properties of which were
carefully studied  with LEPS. The straight section of $l_x =
5.5$ m between the bending magnet D2 and the entrance into the
split pole of LEPS serves as the decay path for $\pi \to \mu x$.
\medskip
\noindent For this setup a beam momentum of $p = 150$ MeV/c has been
chosen, which allows an efficient separation of the beam particles
$(\pi, \mu, e)$ by time of flight (TOF) measurements.
The number of pions at the exit of D2 with a phase
space capable of passing the collimator and reaching LEPS is
determined from  the number of pions detected with LEPS at a
magnetic field setting appropriate for the observation of the direct
beam. The relative beam intensity
is monitored continuously by a ring of
$\mu$-telescopes [4] which are mounted near the position of the
collimator and which register muons emerging from the conventional
pion decay in the beam at the Jacobian angle of about $15^\circ$.
The composition of the beam $(\pi,\mu,e)$ is determined by
TOF measurements between the scintillators at the focal
plane and  the RF of the cyclotron. The reliability of the
calibration of the momentum settings of beamline and LEPS has been
checked to be better than $1\%$ by observation of the direct
beam $(p = 150$ MeV/c) as well as of the muons resulting from forward
and backward (in the CM-system of the pion) decays of
$\pi \to \mu \nu_\mu~(p_\mu = 162$  and
74.2 MeV/c, respectively). Muons originating from $\pi \to \mu x$ are
expected just between these values  at a central momentum of $p_\mu
\approx p_\pi \cdot {m_\mu \over m_\pi} = 113.6$ MeV/c.
\medskip
\noindent At the $\pi E3$ channel the beam for positively charged
particles also contains low-energy protons of high intensity. In
order to avoid severe damage of the intermediate focus detector by
these protons, we have carried out all measurements at this
particular setup with a negatively
charged beam, i.e. searching for a  $\pi^- \to \mu^- \bar{x}$ decay.
\medskip
\noindent In  a second stage of the experiment we installed in
addition superconducting Helmholtz coils at the pivot point of LEPS
(position PP in fig. 1) supplying
a vertical magnetic field up to 2.5 T. This way
beam pions are separated from the muons (originating
from the pion decay in flight) sufficiently in angle, in order to
prevent  them from  entering LEPS. Thus  muons
originating from the pion decay within the  dipole of LEPS do no
longer contribute to the background at the focal plane. In contrast to the
first setup, where we had a complete vacuum beamline between pion
production target and LEPS, in this setup the vacuum systems of the $\pi E3$
channel,
the superconducting magnet and LEPS had to be separated by thin foils.
Upstream and downstream of the superconducting magnet these foils amount
to a thickness of about 32 mg/cm$^2$ each. The energy straggling
due to these foils has been small and tolerable, leading to a
momentum resolution of still better than 0.5\%.
\medskip
\noindent The measurements with this setup have been carried out with
the $\pi^-$ and $\pi^+$ beams of momentum
$p_\pi = 140$ MeV/c and a magnetic field of
about 1.6 T, which gives a deflection angle of $34^\circ$ for pions and
$45^\circ$ for muons, respectively. Slits S1 and S2 have been set to openings
of
44 mm to give maximum beam flux through the collimator of
 10 mm at the position C (fig. 1). Further on a thick copper
collimator with a diameter of 50 mm has been put on the
entrance of LEPS, in order to block beam pions, which are separated
from the muons by the superconducting magnet. In this setup
the decay section for $\pi \to \mu x$ decay is
the straight section between the bending magnet D2 and
the entrance of the superconducting magnet, i.e. $l_x = 2.9$m.
\bigskip
\noindent {\bf 3. Results}
\medskip
\noindent A very valuable feature of LEPS is its intermediate focus
equipped with a 6-plane multiwire proportional chamber, which ---
together with the focal plane drift chamber --- allows a detailed track
reconstruction. Together with the timing information from the
scintillator detector at the focal plane this leads to a
very stringent background reduction in the recorded events. Fig. 2
shows  the results from the measurements with the first setup.
Plotted are the momentum spectra with  cuts on
phase space, TOF and energy loss $(\Delta E)$ informations obtained from the
intermediate and focal plane detectors.
The inset in fig. 2 displays the measurement of
the direct $\pi^-$ beam exhibiting a momentum resolution of better than
$0.3\%$. This measurement also defines the intrinsic line shape as
well as the cuts in the TOF, $\Delta E$
and phase space spectra of the individual LEPS detectors.
\medskip
\noindent  The main part of fig. 2 shows  the $\mu^-$ spectrum in the
momentum bite $p = 107 - 118$ MeV/c, i.e. in the range
where we expect the associately produced muons from the decay $\pi^-
\to \mu^- \bar{x}$
$(p_{\mu} = 112.8 - 114.4$ MeV). Here TOF, $\Delta E$ and phase space cuts
have led
to an overall reduction factor of 30  in the number of events.
\medskip
\noindent These spectra  correspond to an accumulated
number of $1.5 \times 10^9$ pions entering the decay section $l_x = 5.5$m at
the
exit of D2. From the kinematics of the $\pi \to \mu x$  decay we
expect a mainly rectangular shaped peak centered at $p = 113.6$ MeV/c
with a width of 1.5 MeV/c. Folding this peak shape with the
intrinsic line shape as
observed with the direct $\pi$ beam  (fig. 2, inset) produces an
expected
line shape for $\mu$ as shown in fig. 2 (bottom).
We note that the expected line width  is substantially
larger than
the momentum resolution observed with the direct beam.
The measured $\mu$ spectrum  (fig. 2, on the right) does not
exhibit any statistically significant structure of the expected shape,
in particular
also not in the region around $p = 114$ MeV/c.
Hence we  associate the
events registered in the constrained spectrum solely to background
events  arising from conventional $\pi$-decay etc.
For a quantitative analysis of the  spectrum we have first
fitted the background spectrum by a smooth curve (polynomial of
$4^{\rm th}$ to $8^{\rm th}$ degree) and next performed a
$\chi^2$-fit where the expected response $R(x)$ was included and its
amplitude varied to minimize $\chi^2$. As the first two lines of
table 1 show, this inclusion did not improve the fit. The resulting
number of events for $R(x)$ is $195 \pm 111 (1\sigma)$ which
corresponds to an upper limit of the branching ratio of

$$BR(\pi^- \to \mu^- \bar{x}) < 5 \times
10^{-7}~~~~(95\% ~\hbox{C.L.})~~. \eqno(1)$$
The same limit was obtained with the
$\pi^-$-beam using the second setup (lines 3 and 4 of table 1).
\medskip
\noindent
The results of the $\pi^+$-measurements,
performed with the second setup only, are
shown in fig.~3.  Again the $\mu^+$ momentum spectrum is shown with
cuts on phase space, TOF and $\Delta E$ informations from the
intermediate and focal plane detectors. The shape of the
constrained spectrum can be understood from the shape of the
phase space accepted by
LEPS, and is basically due to the circular collimator at the entrance of
LEPS, which selects a bell-shaped momentum-bite of $\Delta p/p =
10\%$. We stress that the same shape of the momentum spectrum is
obtained if we select electrons by a corresponding cut in the TOF
spectrum.
This suggests that the displayed $\mu^+$ spectrum arises
primarily from background such as slit scattering.
\medskip
\noindent At the initial pion momentum $p_\pi = 140$ MeV/c of this run
we expect $x$-particles at a momentum around $p_{\mu} = 106$ MeV/c.
Again no statistically significant structures
are observed. The spectrum corresponds to an accumulated number of
$2 \times 10^{10}$ pions entering the decay path $l_x = 2.9$m at the exit of
D2. As table 1, line 6 shows, the fit to the expected response $R(x)$
yields $- 553 \pm 432$ events. According to the recommendations of
the Particle Data Group [5] only the physical region of positive
values is considered. Then we obtain
as an  upper limit for the branching ratio
$$BR(\pi^+ \to \mu^+ x) < 7 \times 10^{-8} (95\% ~\hbox{C.L.})~~.
\eqno(2)$$
\medskip
\noindent In a recent paper [6] Barger, Phillips and
Sarkar derive limits on the correlation between
branching ratio and lifetime as
given in fig. 4 of  ref. [1] by assuming that $x$ is
a massive neutrino underlying  the standard weak
interaction. From the absence of experimental evidence for
anomalous contributions to $\mu \to e \nu \nu$ and $\pi \to e
\nu$ decays they  find the following limits for the branching ratio
$$2 \times 10^{-8} < BR(\pi \to \mu x) < 8 \times 10^{-5}~~. \eqno(3)$$
Our measurements lower this upper limit by  three orders of
magnitude. Together with the theoretical limits of expression (3)  our
measurements leave only a small region in the branching ratio
for the existence of the $x$-particle.
\bigskip
\noindent {\bf 4. Summary}
\medskip
\noindent Prompted by the reported anomaly in the time spectrum of
single-prong events of KARMEN we have undertaken a first search for
the hypothetical $\pi \to \mu x$ decay utilizing the LEPS-setup at
the $\pi E3$ channel of PSI. This setup is very well suited for
studying the pion decay in flight, in particular if the Q-value of
the decay branch is very small as is the case for the postulated decay
into the $x$-particle. Our first results give a branching ratio of
less than $7 \times 10^{-8}$  at 95\% C.L.
for this decay branch, thus leaving only  a narrow allowed region
in the branching ratio if  combined with the
considerations of Ref. [6] based on  standard weak interactions.
\medskip
\noindent We gratefully  acknowledge valuable discussions with M.
Daum, F. Foroughi, H.-Chr. Walter and B. Zeitnitz. We also would like
to thank the Paul Scherrer Institut for assistance in setting up this
experiment in short time.
\bigskip
\noindent {\bf References}
\medskip
\item{[1]} B. Armbruster et al., Phys. Lett. {\bf B348} (1995) 19
\item{[2]} PSI users' guide (eds.: H.C. Walter, L. Adrian, R. Frosch,
M. Salzmann), CH-5232 Villigen PSI, July 1994, p. 28
\item{[3]} H. Matth\"ay et al., Proc. Int. Symp. on Dynamics of
Collective Phenomena in Nuclear and Subnuclear Long Range
Interactions in Nuclei, ed. P. David (World Scientific, Singapore,
1988), p. 542
\item{[4]} B.M. Barnett et al., Nucl. Instr. Meth. {\bf A297} (1990)
444
\item{[5]} Particle Data Group, Phys. Rev. {\bf D50} (1994) p. 1281
\item{[6]} V. Barger, R.J.N. Phillips and S. Sarkar, Rutherford
Appleton Laboratory, preprint RAL-95-02, March 1995,  Phys. Lett.
{\bf B352} (1995) 365
\vfill\eject
\noindent {\bf Figure Captions:}
\medskip
\itemitem{Fig. 1:} Schematical layout of the  experimental setup.
Perpendicular to the plane of the figure the proton beam hits the
target PT for the production of $\pi, \mu$ and e. These are momentum
and phase space selected in the $\pi E3$ beamline before being
analysed in LEPS. D, Q and H denote dipole, quadrupole and hexapole
magnets in the $\pi E3$
beamline, whereas S$_1$ and S$_2$ denote positions of slits, which
may be varied both in horizontal and vertical directions. At the
position of the last quadrupole magnet of the beamline, a 6
cm long copper collimator (C) with  a diameter  of 10 mm  has been
installed. PP denotes the pivot point of LEPS, where in the second
stage of the experiment superconducting Helmholtz coils have been installed.
3Q denote the quadrupole triplet of LEPS, which
focusses the beam onto the intermediate focus (IF) detector before
it enters the split pole dipole of LEPS. At the focal plane FP a
vertical drift chamber has been installed followed by scintillation
detectors. The straight section between D2 and the entrance into
the dipole of LEPS serves as decay path for $\pi \to \mu x$.
\bigskip
\itemitem{Fig. 2:} Momentum spectra of pions (inset) and muons in
the focal plane of the LEPS spectrometer obtained with a $\pi$ beam
of $T_\pi = 150$ MeV/c and the first setup. The events have been
constrained by $\Delta E$, TOF and phase space information. Abscissa
scales are identical. The solid curve is a $4^{\rm th}$ order
polynomial fit to the data. Also indicated is the expected response
to muons from $\pi^- \to \mu^- \bar{x}$ decay assuming as example a
branching ratio $B = 2 \cdot 10^{-6}$.
\bigskip
\itemitem{Fig. 3:} Momentum spectra of muons in the focal plane of
LEPS obtained with a $\pi^+$ beam of $p_\pi = 140$ MeV/c and the
setup including superconducting Helmholtz coils. The solid curve
represents a $6^{\rm th}$ order polynomial fit to the data. At the
bottom we have indicated the expected response to $\pi^+ \to \mu^+ x$
decay with a branching ratio $B = 2 \cdot 10^{-6}$ as an example.

\vfill\eject
\centerline{Table 1.}
\medskip
\bigskip
\noindent Summary of the statistical analysis of the
measured muon spectra. $p^n$ denotes a polynomial of n$^{\rm th}$
order, $R(x)$ is the expected response function for the $x$-particle,
$DF$ are degrees of freedom in the fit and $N_x$ gives the number of
counts (with $1 \sigma$ uncertainty) in the $x$-region resulting from
the fit. SHC = superconducting Helmholtz coil at pivot point (PP) of
LEPS.
\bigskip
\medskip
$$\offinterlineskip \tabskip=0pt
\vbox{
\halign{\strut
\vrule width 0.8pt # \enskip
&\hfill # \hfill  \enskip
&\vrule #  \enskip
&\hfill # \hfill \enskip
&\vrule # \enskip
&\hfill # \hfill \enskip
&\vrule # \enskip
&\hfill # \hfill \enskip
&\vrule # \enskip
&\hfill # \hfill \enskip
&\vrule # \enskip
&\hfill # \hfill
\tabskip=0pt
&\vrule width 0.8pt # \cr
\noalign{\hrule\hrule}
&&&&&&&&&&&& \cr
&channel & &SHC & &Analysis & &$\chi^2 (DF)$ & &$N_x$ (counts) & &$BR$
(95\%~C.L.) & \cr
&setting & & & & & & & & & & & \cr
&&&&&&&&&&&& \cr
\noalign{\hrule}
&&&&&&&&&&&& \cr
&$\pi^-$ & & & &$p^4$ & &153.4 (143) & &--- & & & \cr
& & &no & & & & & & & & & \cr
&150 MeV/c & & & &$p^4 + R(x)$ & &150.3 (142) & &$195 \pm 111$ & &$< 5.0 \cdot
10^{-7}$ & \cr
&&&&&&&&&&&& \cr
\noalign{\hrule}
&&&&&&&&&&&& \cr
&$\pi^-$ & & & &$p^6$ & &110.8 (102) & &--- & & & \cr
& & &yes & & & & & & & & & \cr
&140 MeV/c & & & &$p^6 + R(x)$ & &109.9 (101) & &$170 \pm 171$ & &$< 4.8 \cdot
10^{-7}$ & \cr
&&&&&&&&&&&& \cr
\noalign{\hrule}
&&&&&&&&&&&& \cr
&$\pi^+$ & & & &$p^6$ & &124.0 (113) & &--- & & & \cr
& & &yes & & & & & & & & & \cr
&140 MeV/c & & & &$p^6 + R(x)$ & &122.4 (112) & &$- 553 \pm 432$ & &$< 7 \cdot
10^{-8}$ & \cr
&&&&&&&&&&&& \cr
\noalign{\hrule\hrule}
}}$$

\vfil
\eject

\centerline{Figure 1.}
\vfill
\epsfxsize=125mm
\epsffile{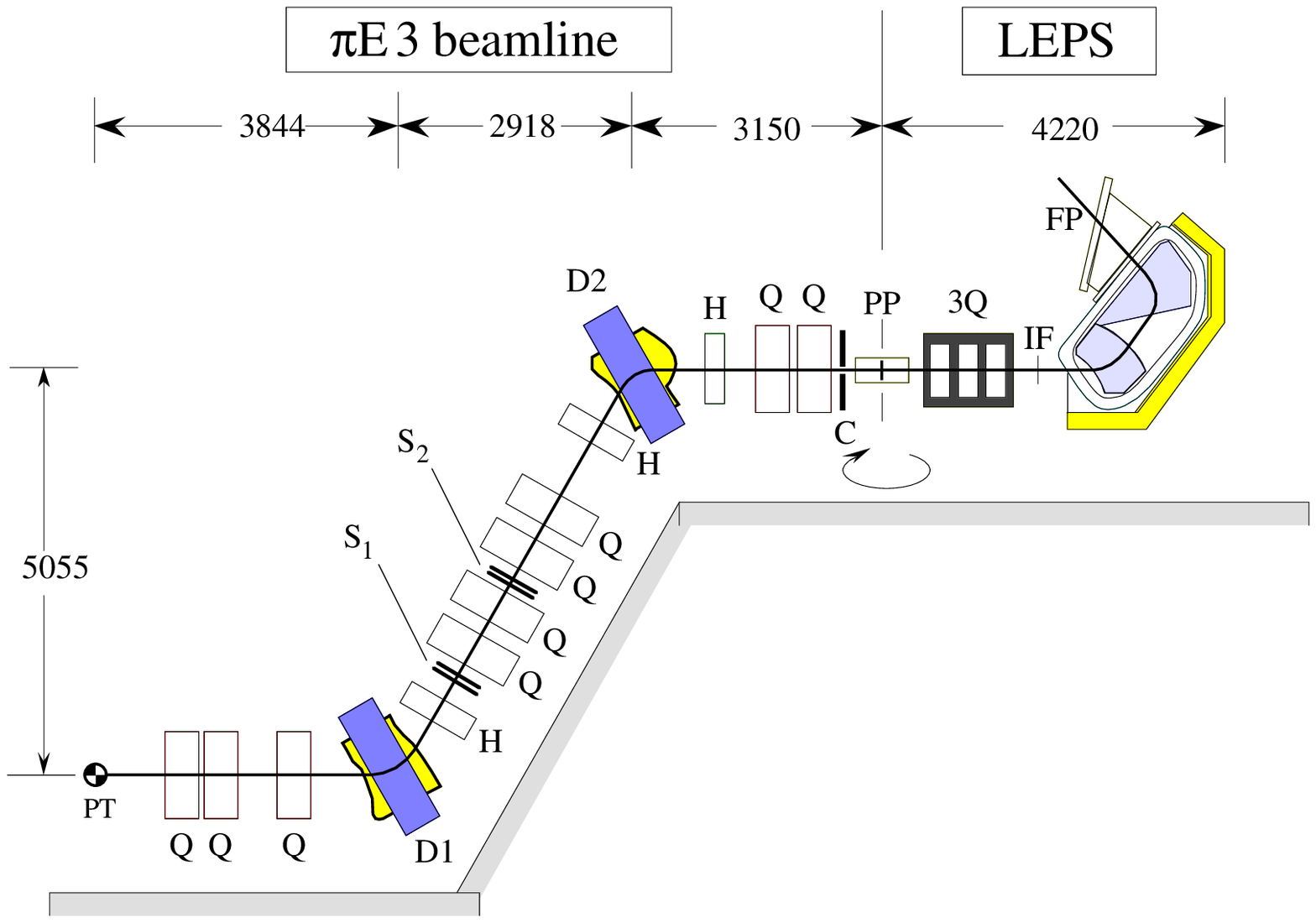}
\vfill
\eject

\vfill
\centerline{Figure 2.}
\vfill
\epsfxsize=125mm
\epsffile{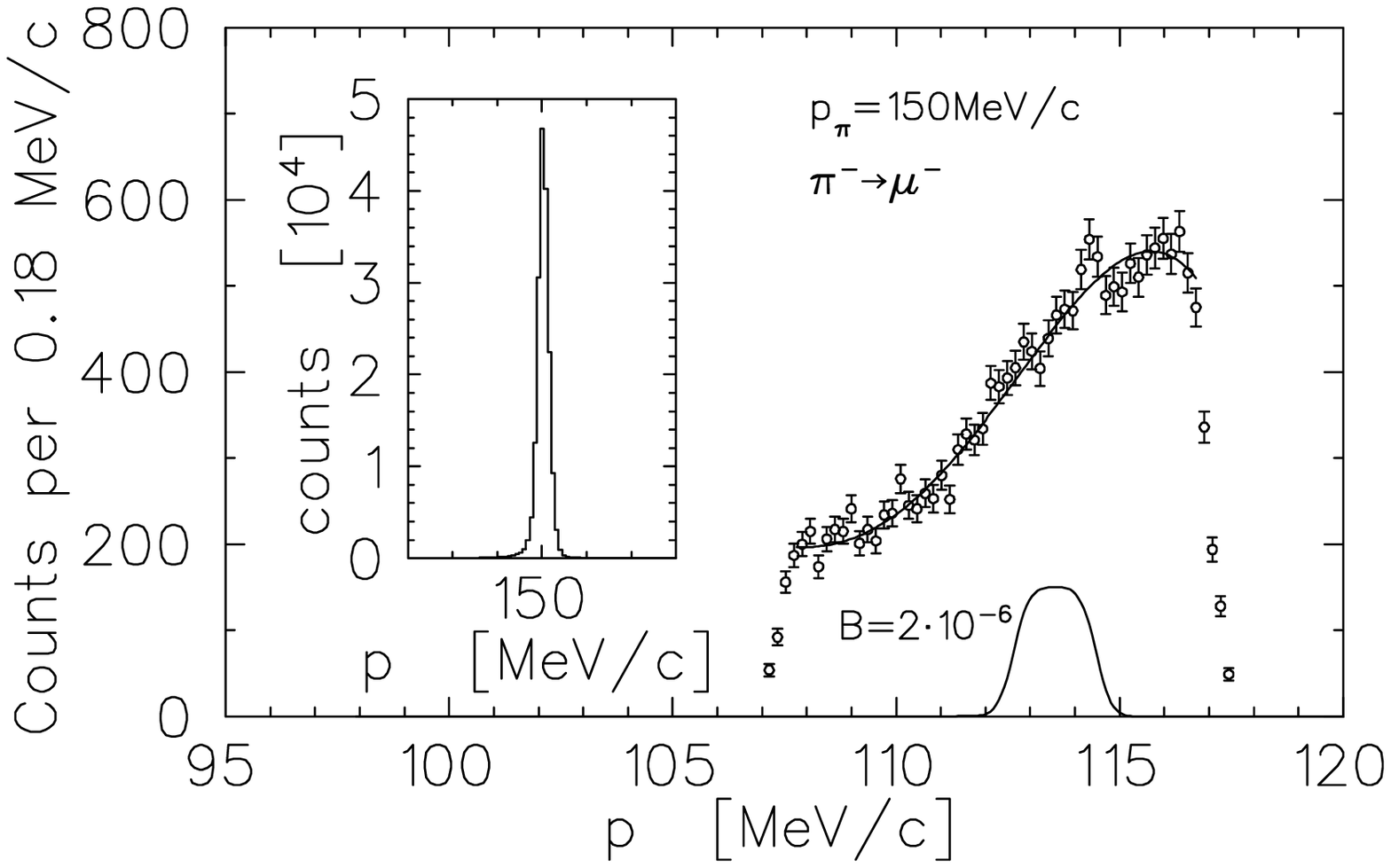}
\vfill
\vfill
\centerline{Figure 3.}
\vfill
\epsfxsize=125mm
\epsffile{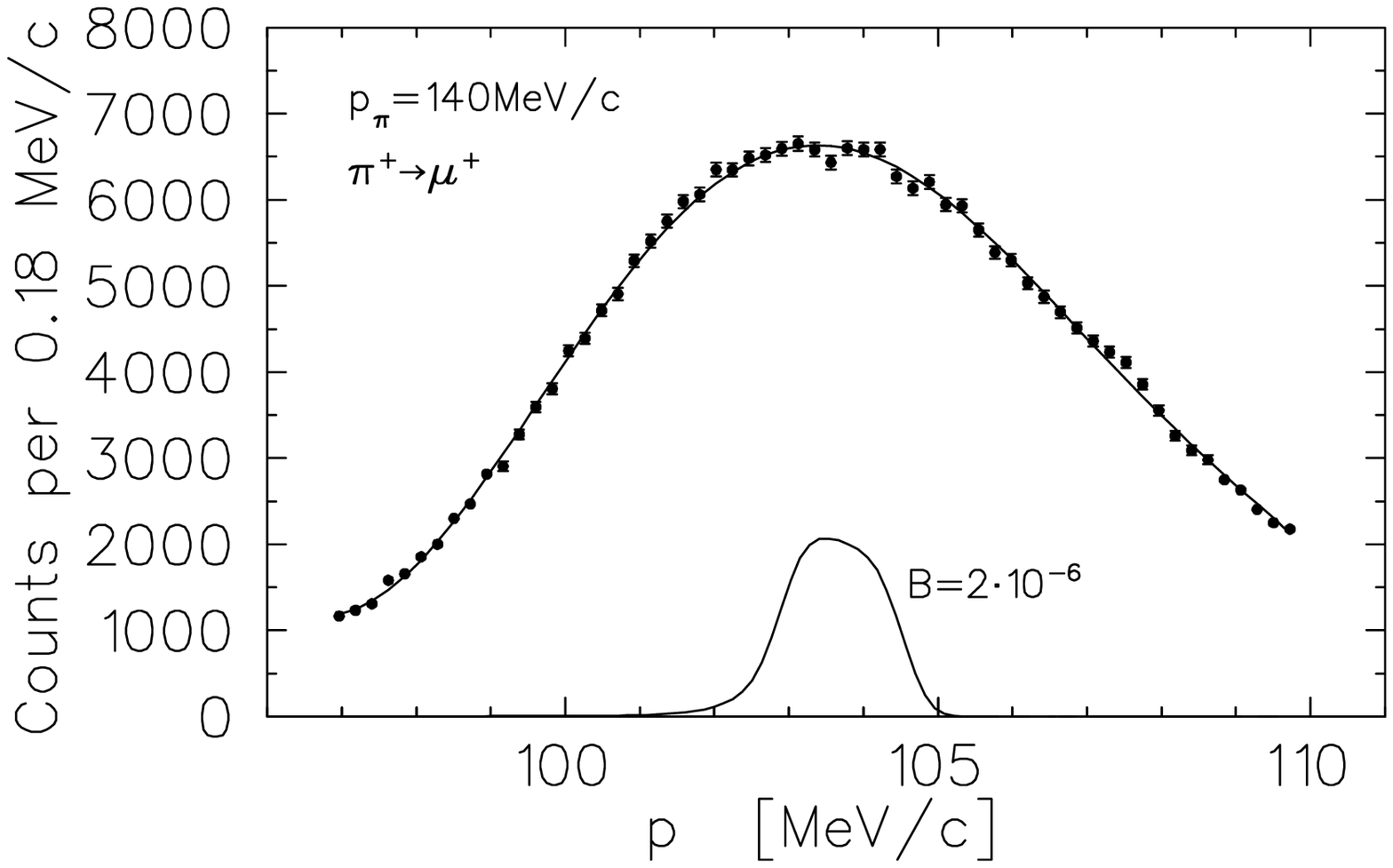}
\vfill

\end